\documentclass[]{article}
\voffset= - 1.0in
\catcode`\@=11
\def\marginnote#1{}
\newcount\hour
\newcount\minute
\newtoks\amorpm
\hour=\time\divide\hour by60
\minute=\time{\multiply\hour by60 \global\advance\minute by-\hour}
\edef\standardtime{{\ifnum\hour<12 \global\amorpm={am}%
        \else\global\amorpm={pm}\advance\hour by-12 \fi
        \ifnum\hour=0 \hour=12 \fi
        \number\hour:\ifnum\minute<10 0\fi\number\minute\the\amorpm}}
\edef\militarytime{\number\hour:\ifnum\minute<10 0\fi\number\minute}
\def\draftlabel#1{{\@bsphack\if@filesw {\let\thepage\relax
   \xdef\@gtempa{\write\@auxout{\string
      \newlabel{#1}{{\@currentlabel}{\thepage}}}}}\@gtempa
   \if@nobreak \ifvmode\nobreak\fi\fi\fi\@esphack}
        \gdef\@eqnlabel{#1}}
\def\@eqnlabel{}
\def\@vacuum{}
\def\draftmarginnote#1{\marginpar{\raggedright\scriptsize\tt#1}}
\def\draft{\oddsidemargin -.5truein
        \def\@oddfoot{\sl preliminary draft \hfil
        \rm\thepage\hfil\sl\today\quad\militarytime}
        \let\@evenfoot\@oddfoot \overfullrule 3pt
        \let\label=\draftlabel
        \let\marginnote=\draftmarginnote
   \def\@eqnnum{(\theequation)\rlap{\kern\marginparsep\tt\@eqnlabel}%
\global\let\@eqnlabel\@vacuum}  }
\makeatletter
\newdimen\normalarrayskip              % skip between lines
\newdimen\minarrayskip                 % minimal skip between lines
\normalarrayskip\baselineskip
\minarrayskip\jot
\newif\ifold             \oldtrue            \def\new{\oldfalse}
\def\arraymode{\ifold\relax\else\displaystyle\fi} % mode of array entries
\def\eqnumphantom{\phantom{(\theequation)}}     % right phantom in eqnarray
\def\@arrayskip{\ifold\baselineskip\z@\lineskip\z@
     \else
     \baselineskip\minarrayskip\lineskip2\minarrayskip\fi}
\def\@arrayclassz{\ifcase \@lastchclass \@acolampacol \or
\@ampacol \or \or \or \@addamp \or
   \@acolampacol \or \@firstampfalse \@acol \fi
\edef\@preamble{\@preamble
  \ifcase \@chnum
     \hfil$\relax\arraymode\@sharp$\hfil
     \or $\relax\arraymode\@sharp$\hfil
     \or \hfil$\relax\arraymode\@sharp$\fi}}
\def\@array[#1]#2{\setbox\@arstrutbox=\hbox{\vrule
     height\arraystretch \ht\strutbox
     depth\arraystretch \dp\strutbox
     width\z@}\@mkpream{#2}\edef\@preamble{\halign
\noexpand\@halignto
\bgroup \tabskip\z@ \@arstrut \@preamble \tabskip\z@ \cr}%
\let\@startpbox\@@startpbox \let\@endpbox\@@endpbox
  \if #1t\vtop \else \if#1b\vbox \else \vcenter \fi\fi
  \bgroup \let\par\relax
  \let\@sharp##\let\protect\relax
  \@arrayskip\@preamble}
\def\eqnarray{\stepcounter{equation}%
              \let\@currentlabel=\theequation
              \global\@eqnswtrue
              \global\@eqcnt\z@
              \tabskip\@centering
              \let\\=\@eqncr
 \halign to \displaywidth\bgroup
    \eqnumphantom\@eqnsel\hskip\@centering
    $\displaystyle \tabskip\z@ {##}$%
    \global\@eqcnt\@ne \hskip 2\arraycolsep
         $\displaystyle\arraymode{##}$\hfil
    \global\@eqcnt\tw@ \hskip 2\arraycolsep
         $\displaystyle\tabskip\z@{##}$\hfil
         \tabskip\@centering
    &{##}\tabskip\z@\cr}
\begingroup\ifx\undefined\newsymbol \else\def\input#1 {\endgroup}\fi
\newfont{\hr}{msbm10}
\newfont{\ams}{msam10}
\font\numbers=cmss12
\font\upright=cmu10 scaled\magstep1
\def\stroke{\vrule height8pt width0.4pt depth-0.1pt}
\def\topfleck{\vrule height8pt width0.5pt depth-5.9pt}
\def\botfleck{\vrule height2pt width0.5pt depth0.1pt}
\def\Zmath{\vcenter{\hbox{\numbers\rlap{\rlap{Z}\kern 0.8pt\topfleck}\kern
2.2pt
                   \rlap Z\kern 6pt\botfleck\kern 1pt}}}
\def\Qmath{\vcenter{\hbox{\upright\rlap{\rlap{Q}\kern
                   3.8pt\stroke}\phantom{Q}}}}
\def\Nmath{\vcenter{\hbox{\upright\rlap{I}\kern 1.7pt N}}}
\def\Cmath{\vcenter{\hbox{\upright\rlap{\rlap{C}\kern
                   3.8pt\stroke}\phantom{C}}}}
\def\Rmath{\vcenter{\hbox{\upright\rlap{I}\kern 1.7pt R}}}
\def\Z{\ifmmode\Zmath\else$\Zmath$\fi}
\def\Q{\ifmmode\Qmath\else$\Qmath$\fi}
\def\N{\ifmmode\Nmath\else$\Nmath$\fi}
\def\C{\ifmmode\Cmath\else$\Cmath$\fi}
\def\R{\ifmmode\Rmath\else$\Rmath$\fi}
\def\Tr{{\rm Tr}}
\def\Im{{\rm Im}}

\def\d{\partial}

\def\2{{1\over 2}}

\def\d{\partial}

\def\bea{\begin{eqnarray}}
\def\eea{\end{eqnarray}}

\def\beq{\begin{equation}}
\def\eeq{\end{equation}}
\def\ba{\beq\new\begin{array}{c}}
\def\ea{\end{array}\eeq}
\def\be{\ba}
\def\ee{\ea}

\begin{document}
%\draft                               %SWITCH ON/OFF DRAFT VERSION%
\begin{flushright}
FIAN/TD-03/01\\
ITEP/TH-16/01
\end{flushright}
\vspace{0.5cm}
\begin{center}
{\LARGE \bf On First-Quantized Fermions in Compact Dimensions
\footnote{Contribution to a special issue of {\em Theoretical and
Mathematical Physics} in honour of 75th birthday of
Vladimir Ya. Fainberg.}}\\
\vspace{0.5cm}
{\Large A.Marshakov
\footnote{e-mail address: mars@lpi.ru,\ mars@gate.itep.ru}}\\
\vspace{0.5cm}
{\it Theory Department, Lebedev Physics Institute, Moscow
~117924, Russia\\ and \\ ITEP, Moscow ~117259, Russia},
\\
\end{center}
\bigskip
\begin{quotation}
We discuss the path integral representation for the fermionic
particles and strings along the lines of \cite{FM1,FM2} and concentrate at
the problems arising when some target-space dimensions are
compact. An example of partition function for fermionic particle at finite
temperature or with one compact target-space dimension is considered in
detail. It is demonstrated that the first-quantized path integral requires,
in general, presence of nonvanishing "Wilson loops" and modulo some common
problems for real fermions in Grassmannian formulation one can try
to reinterpret them in terms of condensates of the world-line
fermions. The properties of corresponding path integrals in string theory
are also discussed.
\end{quotation}

\section{Introduction}
\setcounter{footnote}{0}
\setcounter{equation}{0}

String theory is still believed to be the only possible candidate to the
role of fundamental theory of all interactions though the exact sense of
these words is rather far from being creative. One should better say that
nowadays it
is almost clear that certain problems which are in principle unresolvable in
the context of conventional quantum field theory may be solved within this
more fundamental theory in the future and, for example, the self-consistent
formulation of quantum gravity is among them.

It is also clear that even terminologically {\em string} theory is not very
exact name for this (not already quite) new fundamental theory, since
strings are light excitations only in some part of the whole parameter or
{\em moduli} space of {\em nonperturbative string} (or M-) theory and only
there the theory can be formulated as perturbative expansion in terms of the
Polyakov path integral \cite{Pol81}. Naively this situation is rather similar to quantum
field theory when only at weak coupling it can be presented as (infinite)
set of harmonic oscillators or free particles. However, the very fact that a
theory is formulated as a sum over string world sheets implies completely
different from conventional quantum field theory way of summing the states
propagating along string loops, which basically means that string theory is
{\em not} a quantum field theory. The problem of constructing {\em string field
theory} or second-quantized theory of strings is a closely related issue.

This problem was among main interests of my teacher Vladimir Yakovlevich
Fainberg when he has drawn my attention to this, very new at that time,
branch of theoretical physics. We started to understand very soon that
string theory is essentially first-quantized and that it is strongly related
with the properties of two-dimensional geometry responsible for correct
counting in string loops. The simplest analog of string path integral is
the path integral over particle's world lines and it can be used
for computation of the Green functions of propagating particles, say,
Dirac fermions, in external
backgrounds \cite{FM1}, being a one-dimensional analog of the
Fradkin-Tseytlin effective action \cite{FraTse} in string theory.

The one-dimensional analog of the Polyakov path integral simply results
into the proper-time representation of propagators or partition functions
of point particles, looking thus to be too complicated formulation of a
very simple problem. However, it provides an important geometrical meaning
and it is
interesting to see methodologically how the well-known answers arise if
one starts from one dimensional analog of string and integrate over
co-ordinates of particle, extra variables living on a world-line
(responsible for "internal" degrees of freedom) and one-dimensional
"geometries". Introducing world-line supersymmetry \cite{BDHDZ}
we immediately obtain
space-time fermions, but already in this first nontrivial example the
integral over one-dimensional "supergravity" becomes ill-defined due to
presence of {\em real} world-line fermions, this is a well-known problem,
discussed,
for example in \cite{Pol,Dot,FM1}. The problem can be resolved, however, if
one thinks of a fermionic particle as of a zero-length limit of fermionic
Neveu-Schwarz-Ramond (NSR) string
\footnote{An alternative approach uses the Green-Scwharz or target-space
fermions from the very beginning and leads to the formalism of superparticle
and Green-Schwarz superstring. In contrast to fermionic path integrals,
describing the propagation of fermion only, the superparticle contains both
boson and fermion, respecting space-time supersymmetry.}.

In this note we are going to discuss another side of the same problem which
arises in the case when target-space is not just Minkowski or Euclidean
space ${\bf R}^D$
but has one or several compact dimensions. The simplest example of
this situation -- a partition function at {\em finite temperature} when one of
dimensions is compactified and the target-space is
${\bf R}^{D-1}\times{\bf S}^1$. It will be shown below that this partition
function may be defined "in spirit" of string theory, i.e. starting from
one-dimensional (super)geometry though again we run into problems related to
the existence of real fermions. Nevertheless these problems can be
understood better if one thinks of a point particle as of a "field-theory"
limit of string.

\section{First Quantized Relativistic Bosons and Fermions
\label{ss:prop}}

The path integral approach to the perturbative string theory \cite{Pol81} is
quite nontrivial but very direct generalization of the first-quantized
representation of quantum field theory as world-line integrals over the
trajectories of relativistic particles (and anti-particles). The integral
over embeddings of the world-sheets and/or world-lines should be
supplemented by integration over two-dimensional and/or one-dimensional
geometries (or super-geometries).
In string theory the integral over two-dimensional geometries leads immediately to
nontrivial consequences for the properties of the target space-time, but in
the "model example" of point particles it can be simply reduced to the
finite-dimensional integral over the lengths of world lines or what is
called the integral over Feynman parameters in quantum field theory.

The simplest illustrative example of this technique is given by path
integral representation for the propagator of bosonic spinless particle
\be
G(X_f,X_i) = \int De \int_{X_i}^{X_f} DX e^{-\2\int {\dot{X}^2\over e} + em^2}
\label{prop}
\ee
where integral over one-dimensional geometry is presented by the integral
over (square root of) one dimensional metric $e$ such that proper length of
the world-line is $T = \int_0^1 e(t)dt$. Naive integration in (\ref{prop})
first over $De$
\be
G(X_f,X_i) \sim \int_{X_i}^{X_f} DX e^{-m\int \sqrt{\dot{X}^2}}
\ee
gives rise to the non Gaussian integral over embeddings $X = \{ X_\mu(t)\}$
only
\footnote{This is a naive infinite-dimensional analog of the exact formula
(\ref{int10}) to be used below.} so that the most effective way to deal with
integrals like (\ref{prop}) is the opposite -- to integrate first over
embeddings $X$ (with boundary conditions $X(0)=X_i$ and $X(1)=X_f$)
in the Gaussian form (\ref{prop}) and to use explicit co-ordinates
in the {\em moduli} space of one-dimensional geometries (metrics factorized
over reparameterizations of world lines) to construct measure $De$.

This is rather trivial procedure, which can be thought of as a variant
of the Faddeev-Popov trick, or, even more simple -- a generalization of
the finite-dimensional formula of Gaussian integration, the details can be
found, for example, in \cite{CMNP,Dot,FM1}. The pretty simple and well-known
result is that one can choose a "gauge"
\footnote{The terminology "gauge" is not, in fact, very strict here since we
are integrating over the whole space of metrics on world-line after choosing
appropriate co-ordinates. However, we will keep this word throught the paper
since this is already commongly accepted terminology and since the integral
over all metrics is equivalent to the integral over their "classes of
equivalence" or lengths $T$.}
$e(t)=T=const$ ($T$ is the length of a
path) and rewrite
integration over $De$ as an integral over reparameterizations and
(one-dimensional) integral over the lengths $T$. The first one, if measure $De$
is normalized by dividing onto the volume of reparameterization group, is
trivial (nothing depends on reparameterizations of world-lines!) and gives
unity, so that, finally, for open trajectories one gets
\be
\int De \int_{X_i}^{X_f} DX e^{-\2\int {\dot{X}^2\over e} + em^2} =
\int_0^\infty dT \int_{X_i}^{X_f} DX e^{-\2\int {\dot{X}^2\over T} + Tm^2}
\ee
and performing (now gaussian!) integration over $X$'s we finally obtain
a well-known formula \cite{CMNP,FM1,FM3}
\be
\label{scaprop}
\int_0^\infty dT e^{-\2 Tm^2-{(X_f-X_i)^2\over 2T}}\det \left(-{1\over T^2}
{d^2\over dt^2}\right)^{-D/2} =
\int{ d^Dp\over (2\pi)^{D/2}}\ e^{ip\cdot(X_f-X_i)}{1\over p^2 + m^2}
\ee
where $D$ is dimension of space-time with (Euclidean) co-ordinates $X_\mu$,
$\mu = 1,\dots,D$.

This pretty simple logic can be generalized to the first-quantized fermions.
Indeed, introducing supersymmetry on world line, one immediately comes to
the space-time fermion \cite{BDHDZ}, for example in the above sense --
computation of the world-line integral with fixed ends gives the Dirac
propagator.

The corresponding world-line action may be determined by invariancy under
world-line supersymmetry transformations
\be
\delta X = \epsilon\Psi
\\
\delta\Psi = -\epsilon\left({\dot X} + \2\chi\Psi\right)e^{-1}
\\
\delta\chi = - 2{\dot\epsilon}
\\
\delta e = - \epsilon\chi
\label{susy}
\ee
and contains Grassmann fermionic variables $\Psi_\mu$, which after quantization
become gamma-matrices \cite{BDHDZ}, so that the wave function carries now
also a {\em space-time} fermionic index, since it is a vector in
representation of the Clifford algebra.

The path integral representation for the Dirac particle propagator
\cite{Pol,Dot,FM1} has a similar form to (\ref{prop})
\be
G(X_f,X_i) = \int De~D\chi~\int_{X_i}^{X_f}DX\int~D\Psi~
e^{-\2\int_{dt}{\dot{X}^2\over e} + \Psi{\dot\Psi} +
{\chi\over e}\Psi{\dot X} + m^2(e + {1\over 4}\chi d_t^{-1}\chi)}
\label{ferprop}
\ee
where one adds the integration over the Grassmann variables $\Psi_\mu$ as well
as the integral over one-dimensional geometries
$\int De$ is replaced by the integral over one-dimensional
super-geometries $\int De~D\chi$. The path integral (\ref{ferprop})
has the following properties (investigated in detail in \cite{FM1}):
\begin{itemize}

\item The bosonic part is the same as in (\ref{prop}) giving rise to a
contribution to the integral identical to l.h.s. of (\ref{scaprop}).

\item The fermionic integral is ill-defined since we deal with {\em real}
fermions $\Psi_\mu$ and $\chi$.

\item Real first-order fermion $\Psi_\mu$ coincides with its own momentum
${\delta S\over\delta\dot\Psi_\mu}$. That
leads to appearance of the Clifford algebra after quantization. In the path
integral (\ref{ferprop}) it means that we integrate around the classical
trajectory $\Psi_\mu = \gamma_\mu$ with $\gamma_\mu$ to be replace by Dirac
gamma-matrices $[\gamma_\mu,\gamma_\nu]_+=\delta_{\mu\nu}$.
More strictly the path integral (\ref{ferprop}) defines the
{\em symbol} of the propagator (it is especially clear if one considers
sample computations in external fields, see \cite{FM1,FM3} for details).

\item The Gaussian integral over the fluctuations $(\Psi_\mu-\gamma_\mu)$
gives rise to the determinant of the first-order differential operator
$\det\left({d\over dt}\right)^{D/2}$ which does not depends on metric $e$ on
world line or its length $T$. This can be established in several ways, for
example, one may compute it via "discretization". Another method of
computation the same determinant is to assign to the fluctuations the
{\em antiperiodic} boundary
conditions at the end of the world line.

\item Like "metric" $e$ the "gravitino" $\chi$ by local transformations
can be "gauged" to the form $\chi(t) = {\cal X} = const$, where ${\cal X}$ is
now a {\em Grassmannian} constant. The Grassmann
integral over ${\cal X}$ gives rise to the most essential contribution into
Dirac propagator
\be
\label{dirrop}
\int_0^\infty dT\ {e^{-\2 Tm^2-{(X_f-X_i)^2\over 2T}}\over\det
\left(-{1\over T^2}
{d^2\over dt^2}\right)^{D/2}}\int d{\cal X} e^{\ {{\cal X}\over 2T}
\left(\gamma\cdot(X_f-X_i)+ m\gamma_5\right)} = \\ =
\int_0^\infty {dT\over T^{1+D/2}}\ e^{-\2 Tm^2-{(X_f-X_i)^2\over 2T}}
{1\over 2T}\left(\gamma\cdot(X_f-X_i)+ m\gamma_5\right) =
\int{ d^Dp\over (2\pi)^{D/2}}\ e^{ip\cdot(X_f-X_i)}\
{\gamma\cdot p + m\gamma_5\over p^2 + m^2}
\ee
where $\gamma_5$ (to be replaced by product of all Dirac matrices
$\gamma_\mu$) can be thought of \cite{Pol} as a classical value of auxiliary
Grassmann variable $\Psi_5$ appearing when one gets rid of nonlocal term in
the action in (\ref{ferprop}).

\item The "illness" of real fermions is seen when one derives the
integration measure over $d{\cal X}$ from the "first principles" of
world-line theory. To do this, one needs, for example, to {\em complexify}
the field $\chi$ \cite{Dot,FM1}. The only reason for doing this, known to the
author, is if one says that the propagator for Dirac particle is a limiting
case for the propagator of the open Ramond string \cite{FM2,FM3}. In string
theory world{\em sheet} supergeometry has two gravitino fields $\chi$ and
$\bar\chi$ (see sect.~\ref{ss:string} below) and the integration measure can
be defined rigorously.
\end{itemize}

Apart for the problems with real fermions listed above, the first-quantized
technique works quite effective both for bosons and fermions. One can
illustrate this, for example, considering generating functionals for
particles in external fields (see details and references in \cite{FM3})
-- the one-dimensional analogs of the Fradkin-Tseytlin string
effective actions. One adds, for example, exponent of the interaction term with
the gauge field
\be
\label{gauge}
\int_{dt} \left({\dot X}_\mu(t) A_\mu(X(t)) +
{e(t)\over 2}F_{\mu\nu}(X(t))\Psi_\mu(t)\Psi_\nu(t)\right)
\ee
(ordered $P$-exponent in the non-Abelian case) and this generating
functional gives rise to all desired results.

In what follows we will see that {\em background} terms and especially
those exactly given by (\ref{gauge}) are extremely important for the theory
with compact dimensions. The interaction of string with gauge field
is given by the same integral (\ref{gauge}) which is taken, in this case,
over the one-dimensional boundary of two-dimensional world-sheet.
In the first-quantized ideology of string theory
one starts with some {\em arbitrary} background which is then
"tuned" by propagating string. If we think of a particle being the
zero-length limit of string, the same ideology may be applied directly
in the case of integration over world-lines.

In general, exponent of (\ref{gauge}) corresponds to nontrivial external
gauge field. For vanishing field strength $F_{\mu\nu}=0$, $A_\mu$ is a pure
gauge and the first term in (\ref{gauge}) can be integrated out. This is not
the case, however, if one allows compact dimensions, then  vanishing
$F_{\mu\nu}$ still allows nonvanishing "Wilson loops" and
the path integral is defined, up to exponent of some phase
\be
\label{phase}
\int_{dt} {\dot X}_\mu(t) A_\mu(X(t)) = \oint A_\mu dX_\mu
\ee
For the flat target space one may always choose $A_\mu=const$ and the nonzero phase
can be "generated" only for the compact direction, say in the case of
${\bf R}^{D-1}\times{\bf S}^1$ only $A_0$ may be nonvanishing (index "$0$" here
and below would correspond to the compact part of
${\bf R}^{D-1}\times{\bf S}^1$). Possible arising of nontrivial
phase for the first-quantized path integral will be essentially used below
and one may even try to interpret this phase as coming from "internal"
degrees of freedom, living on world-lines and world-sheets.

\section{Free Energy of Point Particles}

Let us now turn more closely to the problem we are going to discuss below in detail.
In contrast to sect.~\ref{ss:prop} consider {\em closed}
trajectories, corresponding to the 1-loop partition functions, which from
the point of view of quantum field theory are equal to
\be
\pm\2\log\det (p^2 + m^2)
\label{1loopft}
\ee
where two different signs correspond either to fermions or to bosons. These
determinants have well-known proper-time representations, for example, in bosonic
case one can write
\be
- \log\det (p^2 + m^2) = - \Tr\log (p^2 + m^2) =
- \int{d^Dp\over(2\pi)^D}\ \log (p^2 + m^2) =
\\ =
\int_0^{\infty} {dT\over T}\int{d^Dp\over(2\pi)^D}\ e^{-\2 T(p^2 + m^2)} =
{1\over (2\pi)^{D/2}}\int_0^{\infty} {dT\over T^{1+{D\over 2}}}e^{-\2 T m^2}
\label{1lcomp}
\ee
The last expression can be considered as a similar to (\ref{prop})
(first-quantized) path integral representation for a particle,
\be
\int DeDX e^{-\2\int {\dot{X}^2\over e} + em^2}
\label{pathint}
\ee
but now on {\em periodic} trajectories i.e. with boundary conditions $X(0) = X(1)$
\footnote{$X_f=X_i$ with extra integration over this point
if compare to the integral over co-ordinates in (\ref{prop}).}.
The computation can be
performed in a standard way, again choosing the "gauge"
$e(t) = T = \int_0^1 e(t)dt$, when it is reduced to
\be
\int_0^{\infty}{dT\over T}e^{-\2 Tm^2}\int DX e^{-\int {\dot{X}^2\over 2T}}
\ee
since the integration measure over the metric $De$ on world-line, induced
by $\| \delta e\|^2 = \int_{dt} {(\delta e)^2\over e}$, gives rise on
{\em periodic} trajectories to $\int_0^{\infty}{dT\over T}$ (in contrast to
$\int_0^{\infty}dT$ on {\em open} world lines).
The integral over $DX$ separates into the zero-mode part $dX^{(0)}$
which gives the contribution $\left(T\over 2\pi\right)^{D\over 2}$,
coming from the reparameterization invariant definition of the integration
measure over zero modes,
times the volume of target space ${\rm Vol}({\bf R}^D)$,
while the rest is again
$ \det\left( -{1\over T^2}{d^2\over dt^2}\right)^{-{D\over 2}}$, to be
computed in a standard way. For periodic boundary conditions
\be
\log\det\left( -{1\over T^2}{d^2\over dt^2}\right) =
4\sum_{n=1}^{\infty}\log{2\pi n\over T} = 2\log T + {\rm const}
\ee
since we have double degenerated eigenvalues
$\left({2\pi n\over T}\right)^2$ with $n=1,2,\dots$. Altogether, after
normalization to the volume of target space ${\rm Vol}({\bf R}^D)$,
it gives the last expression in (\ref{1lcomp}).

If now one considers a particle in space-time with compact directions (for
example the so called theories at finite temperature when
$X_0 \sim X_0 + \beta$), one should make a substitution
$\int {dp_0\over 2\pi} \to {1\over \beta}\sum_{n\in\bf Z}$
when computing the trace in (\ref{1lcomp}), so that
\be
\2\int{d^D p\over(2\pi)^D}\ \log (p^2 + m^2)\rightarrow
{1\over 2\beta}\sum_{n\in\bf Z}\int_{d\bf p}\log ({\bf p}^2 + m^2 +
{4\pi^2n^2\over\beta^2}) =
\\ =
{1\over\beta}\int_{d\bf p}\log\sinh{\beta\over 2}\sqrt{{\bf p}^2 + m^2} =
{1\over\beta}\int_{d\bf p}\log\left(1 - e^{-\beta\omega_{\bf p}}\right)
\label{1ltc}
\ee
up to a divergent constant, including the vacuum energy of the infinite
system of oscillators $\int_{d\bf p}{\omega_{\bf p}\over 2}$,
where $\omega_{\bf p} = \sqrt{{\bf p}^2 + m^2}$ and $d{\bf p} \equiv
{d^{D-1}p\over(2\pi)^{D-1}}$ stays for the integration only over the "space"
momenta. In (\ref{1ltc}) we have used that
\be
\sinh\pi x = \pi x\prod_{n=1}^{\infty}\left( 1 + {x^2\over n^2}\right)
\ee
Now the path integral representation (\ref{pathint}) gives for this case
\be
\int_0^{\infty}{dT\over T}e^{-\2 Tm^2}\int DX e^{-\int {\dot{X}^2\over 2T}} =
{{\rm Vol}({\bf R}^{D-1})\over(2\pi)^{D-1\over 2}}\cdot
{\beta\over\sqrt{2\pi}}\int_0^{\infty}{dT\over T^{1+{D\over 2}}}e^{-\2 Tm^2}
\sum_{n\in{\bf Z}}e^{-{\beta^2n^2\over 2T}}
\ee
which includes the $\theta$-function sum over "wrappings" along the compact
$X_0$-direction
\be
\label{thetab}
\theta\left(0 \left| {i\beta^2\over 2\pi T}\right.\right) =
\sum_{n\in{\bf Z}} e^{-{\beta^2n^2\over 2T}}
\ee
and the extra
factor $\beta \sim \int dX_0^{(0)}$ comes from the integration over
the zero mode in compact direction
(${\rm Vol}({\bf R}^D)\to\beta\cdot{\rm Vol}({\bf R}^{D-1})$). This sum
can be interpreted as a sum over ensemble of particle (each term corresponds
to $n$ particles, propagating in the loop), thus giving rise to statistic
quantity in terms of formally one-particle, first-quantized integral.
The result, normalized now onto the volume of "space-dimensional" part
${\rm Vol}({\bf R}^{D-1})$ only, is free energy $\beta{\cal F}$, since
\be
{\beta\over\sqrt{2\pi}}\sum_{n=1}^{\infty}
\int_0^{\infty}{dT\over T^{3\over 2}}{e^{-\2 Tm^2}\over
T^{D-1\over 2}}e^{-{\beta^2n^2\over 2T}} =
{\beta\over\sqrt{2\pi}}\sum_{n=1}^{\infty}\int_{d\bf p}
\int_0^{\infty}{dT\over T^{3\over 2}}
e^{-\2 T({\bf p}^2+m^2) + {\beta^2n^2\over 2T}} =
\\
= \beta\int_{d\bf p}\sum_{n=1}^{\infty}{e^{-\beta n\omega_{\bf p}}\over\beta n} =
- \int_{d\bf p}\log\left( 1 - e^{-\beta\omega_{\bf p}}\right) = \beta{\cal F}
= \log Z
\label{bose}
\ee
In order to present the last computation one
should use the integration formula
\be
\label{int10}
\int_0^{\infty}{dT\over\sqrt{T}}e^{- a^2T - {b^2\over T}} =
 {\sqrt{\pi}\over |a|}e^{-2|ab|}
\ee
which is literally used in (\ref{bose}) after a "modular" transformation
$T\to{1\over T}$.
Thus, we have checked that in bosonic case the first-quantized formalism
immediately, without any subtleties, like introducing nontrivial backgrounds,
leads to correct well-known result.

\section{The path integral for fermions
\label{ss:fermi}}

However, the situation is not as simple
in the fermionic case, when one should get instead of (\ref{bose})
\footnote{Notice, that the first-quantized path integral gives the {\em
logarithm} of the partition function (free energy) so that in both bosonic
(\ref{1ltc}) and fermionic (\ref{1lf}) cases one gets the sums of infinitely
many terms, corresponding to the expansion of logarithms. If the result of
the first-quantized computation were the partition functions themselves
(without logarithms), then in bosonic case one would still see the sum of
infinitely many terms (geometric progression) while in the fermionic case
there would be only two contributions of two-level system.}
\be
\label{1lf}
- {1\over\beta}\int_{d\bf p}\log\left( 1 + e^{-\beta\omega_{\bf p}}\right)
= \int_{d\bf p}\sum_{n=1}^{\infty}{e^{-\beta n\omega_{\bf p}}\over\beta n}
e^{i\pi n} \sim
\int_0^{\infty}{dT\over T^{1+{D\over 2}}}e^{-\2 Tm^2}
\sum_{n\in{\bf Z}}e^{-{\beta^2n^2\over 2T} + i\pi n}
\ee
where the last factor
\be
\label{thetaf}
\theta\left({1\over 2} \left| {i\beta^2\over 2\pi T}\right.\right) =
\sum_{n\in{\bf Z}} e^{-{\beta^2n^2\over 2T}+i\pi n}
\ee
is the only essential contribution of the world-sheet fermions.
In conventional terms of quantum field theory, or better
{\em space-time} fermions, this effect comes from well-known shift to
the Matsubara half-integer frequencies. But let us stress here that this has
nothing in common with
our discussion based on {\em world-line} fermions and below in this section
we are going to look at this problem from the perspective of first-quantized
formalism.

The fermionic analog of the path integral (\ref{pathint}) has the form
\be
\int De~D\chi~DX~D\Psi~e^{-\2\int_{dt}{\dot{X}^2\over e} + \Psi{\dot\Psi} +
{\chi\over e}\Psi{\dot X} + m^2(e + {1\over 4}\chi d_t^{-1}\chi)}
\label{pathfer}
\ee
where integral over $DX$ is again considered over closed trajectories
$X(1)=X(0)$. As in the case of propagators considered in
sect.~\ref{ss:prop}, the integration is almost identical to the bosonic case
and gives
\be
\int_0^{\infty}{dT\over T}e^{-\2 Tm^2}
\int DX~e^{-\2\int_{dt}{\dot{X}^2\over T}}
\int D\Psi~e^{-\2\int_{dt} \Psi{\dot\Psi}}
\int D\chi~e^{-\int_{dt}{1\over 8}\chi d_t^{-1}\chi}
e^{-\int_{dt} {1\over 2T}\chi\Psi{\dot X}} =
\\
= {{\rm Vol}({\bf R}^{D-1})\over(2\pi)^{D-1\over 2}}\cdot
{\beta\over\sqrt{2\pi}}\int_0^{\infty}{dT\over T}e^{-\2 Tm^2}
{1\over T^{D\over 2}}
\left(\det{d_t}^{D\over 2}\right)
\left(\det{d_t}^{-\2}\right)e^{\oint A_{\mu}dX^{\mu}}
\label{fercomp}
\ee
The only essential factor, which distinguishes (\ref{fercomp}) from the
bosonic case is the last factor of the type (\ref{phase}) and which can be
even thought of as coming from fermionic degrees of freedom
$\oint A_{\mu}dX^{\mu} = \int_{dt}{1\over 2T}\chi\Psi{\dot X}$, having
exactly the meaning of a "$\theta $-term" or interaction with gauge field
(\ref{gauge}) $A_\mu = - {1\over 2T}\chi\Psi_\mu$
with vanishing field strength $F_{\mu\nu}(A)=0$ as we
discussed at the end of sect.~\ref{ss:prop}. Let us stress again that such
factor can be the only source for "fermionic anomaly" since the
fermionic determinant $\det{d_t}$ does not depends on metric and its extra
power in (\ref{fercomp}) is absolutely inessential.
Such gauge potential in the example with target-space ${\bf R}^{D-1}\times
{\bf S}^1$ corresponds to a nontrivial loop in
compact direction $X_0\sim X_0 + \beta $, and it means that in our case
one may consider only $A_0 = - {1\over 2T}\chi\Psi_0$ being nonzero.
The bosonic
part of the computation is identical to that one considered above while
two extra determinants in (\ref{fercomp}) -- the degrees of $\det d_t$ --
correspond to the integration over
$D\Psi $ and $D\chi $ and, thus, are independent on metric $e(t) = T$ and/or
should be computed with antiperiodic boundary conditions.

Let us now turn directly to the only possible classical and
"fermion-dependent" contribution
\be
\label{loop}
\oint A_{\mu} dX^{\mu} = - \int_{dt}{\chi\Psi_0\over 2T}{\dot X}_0
\ee
coming from the loop in "temperature" $X_0$-direction. First, already from
general physical
arguments it is clear that the only self-consistent value for this
"background loop" is $A_0\ = {2\pi i\over\beta}(K+\2) $ with any
integer $K\in{\bf Z}$. The corresponding "gauge" symmetry
is a (discrete) ${\bf Z}_2$-symmetry of changing the fermionic sign and
thus one may interpret $A_{\mu} = - {\chi\over
e}\Psi_{\mu}$ just as a "Wilson-line" background (\ref{phase}). In particular, it is
supersymmetric, since
\be
\delta\left({\chi\over e}\Psi_{\mu}\right) = -{d\over dt}
\left({2\epsilon\Psi_{\mu}\over e}\right) + {\epsilon\over e}\left( 2\dot{\Psi}_\mu -
{\chi\over e}\dot{X}_{\mu}\right)
\label{varback}
\ee
where the first term is total derivative and vanishes on periodic trajectories
(even if $\Psi$ is antiperiodic, $\epsilon$ is also antiperiodic and still
$\epsilon\Psi$ is periodic)
while the second is proportional to the equations of motion and vanishes on
classical trajectories.

So, in short, there are few things to be fixed for the computation
(\ref{fercomp}) of the fermionic path integral (\ref{pathfer}):
\begin{itemize}
\item The integration over fluctuations of world-line fermionic fields
$\Psi_\mu$ gives {\em no} nontrivial contribution to the answer for the
temperature partition function, unlike the "Matsubara" target-space
fermions. In more conventional string theory terms this is an extra difference
between the world-sheet Neveu-Schwarz-Ramond and target-space Green-Schwarz
fermions.
\item The only essential fermionic factor, which can arise, has the form of
(\ref{phase})
\be
e^{\oint A_\mu dX_\mu}
\ee
with very special imaginary
\be
\label{a0}
A_\mu = \delta_{\mu,0}{2\pi i\over\beta}(K+\2)
\ \ \ \ \ \
K\in {\bf Z}
\ee
as usual taking values in the "dual torus" -- a circumference with radius
proportional to the {\em inverse} radius of the compact dimension.

\item This "vacuum expectation value" can be thought of as coming from
world line fermions $A_\mu = - {\chi\over 2e}\Psi_\mu$. This is a very
subtle point, since even on classical trajectories if $\langle\chi\rangle =
{\cal X}\neq 0$ one gets $\psi_\mu \sim {\cal X}$ and due to Grassmannian
nature of fermionic variables their product vanishes,
{\em but} $\langle\chi\Psi_\mu\rangle = A_\mu\neq 0$, with $A_\mu$ defined by
(\ref{a0}). The naive inconsistency of this
setting is related, again, to the problem of real fermions (the Grassmann
algebra is one-dimensional) we are coming
back all the time. Below, in sect.~\ref{ss:string}, we consider this
problem in "regularized" setup of string theory, looking at the fermionic
particle integral as at the limit of path integral for the fermionic string.

\item On more rough level, one may even forget about all complications with
the fermionic variables. For example, just say, that {\em any}
first-quantized path integral in space with compact directions is defined up
to the factors like (\ref{phase}) with $F_{\mu\nu}=0$ but without any other
restrictions to $A_\mu$. It means, in particular, that the partition
function will get a contribution
\be
\label{thetagen}
\theta\left(\alpha \left| {i\beta^2\over 2\pi T}\right.\right) =
\sum_{n\in{\bf Z}} e^{-{\beta^2n^2\over 2T}+2\pi in\alpha}
\ee
of which (\ref{thetab}) and (\ref{thetaf}) are particular cases. The generic
factor (\ref{thetagen}) would correspond to {\em anyon}, a particle with
fractional statistics determined by arbitrary phase
$\alpha\in {\bf R}/{\bf Z}$. If we consider, however, a simply-connected
target-space (like ${\bf R}^{D-1}$), and wave-function depends only on {\em
position} in this target-space, or co-ordinate $X$, and does {\em not}
"remember" the path -- by standard "Landau argument" the only choices for
the phase are $\alpha=0,\2$, so that from (\ref{thetagen}) we come back
either to (\ref{thetab}) or to (\ref{thetaf}).
\end{itemize}

So, we have tried to argue in this section that the careful treatment of the fermionic
world-line integral gives rise to expected result in the case of compact
target-space as well as in the case of usual Minkowski/Euclidean space.
However, we again run into a problem with the (integration over) world-line
real fermions which has no rigorous resolution without complexification or
"raising" the problem to the string level.

\section{The fermionic string path integral
\label{ss:string}}

Let us, finally, consider the fermionic string path integral \cite{Pol81}
\be
\label{fstring}
  \int Dg~D\chi~DX~D\Psi~\exp\left(-{1\over 2\pi\alpha'}
\int_\Sigma\bar{\partial}X\partial X+
  \Psi\bar{\partial}\Psi + \bar{\Psi}\partial\bar{\Psi} +
\chi\Psi\bar{\partial}X + \bar{\chi}\bar{\Psi}\partial X +
\2\bar{\chi}\chi\bar{\Psi}\Psi\right)
\ee
where action now is invariant under local two-dimensional supersymmetry
\cite{BDH&DZ}.
In contrast to the particle case (\ref{pathfer}) action in (\ref{fstring})
contains now {\em two} cubic "complex conjugated" terms and {\em quartic}
term $\bar\chi\chi\bar\Psi_\mu\Psi_\mu$, containing both fermions and both
gravitinos. In the case of open fermionic string one should impose boundary
conditions $\Psi = \pm\bar\Psi$, $\chi = \pm\bar\chi$ on the free boundaries
of the world-sheet, and for the Ramond sector (the same sign at all
components of the boundary)  one gets in
the limit $\alpha'\to 0$ the Dirac particle. Due to existence of both $\Psi$
and $\bar\Psi$ and two gravitinos there is no problem any more with
constructing measure for the fermionic string.

This is a kind of "two-dimensional regularization" of one-dimensional
Grassmann algebra considered in sect.~\ref{ss:fermi}. Only at the boundaries
of the world-sheet two independent generators become connected by the
boundary conditions, but now requirement that $\langle\chi\Psi\rangle\neq 0$
(and $\langle\bar\chi\bar\Psi\rangle\neq 0$) is absolutely consistent in the
"interior" of $\Sigma$. Reparameterization invariance requires only for
these quantities to be (linear combinations of) holomorphic or
antiholomorphic differentials, i.e.
existence of "vacuum averages"
\be
\langle\chi\Psi_\mu\rangle = \sum_k a_\mu^{(k)}d\omega_k
\ee
and
\be
\langle\bar\chi\bar\Psi_\mu\rangle = \sum_k \bar a_\mu^{(k)}d\bar\omega_k
\ee
where we have chosen $\{ d\omega_k\}$ to be the set of canonical
differentials on $\Sigma$, $\oint_{A_j}d\omega_k=\delta_{jk}$,
$j,k=1,\dots,g={\rm genus}(\Sigma)$ and $\{ d\bar\omega_k\}$ -- their complex
conjugated.

In the Ramond sector, which directly "regularizes" the fermionic particle
case, one can choose $\langle\chi\rangle = {\cal X}$ and
$\langle\bar\chi\rangle = \bar{\cal X}$, with ${\cal X} = \pm \bar{\cal X}$
on the boundaries of the world-sheet, so that quartic term vanishes and
the normalization conditions come only from the {\em cubic} terms in
(\ref{fstring}), giving
\be
\label{norm}
\int_\Sigma \chi\Psi\bar\d X = \sum a_\mu^{(k)}R_\mu^{(j)}(\Im T)^{-1}_{ij}
\int_\sigma d\omega_k\wedge d\bar\omega_i = \sum_k a_\mu^{(k)}R_\mu^{(k)}
= {\rm (half)\ integer}
\ee
together with complex conjugated. In (\ref{norm}) we have introduced
$T_{ij} = \oint _{B_i}d\omega_j$ -- the period matrix of $\Sigma$.

Relation (\ref{norm}) is a direct two-dimensional generalization of
(\ref{loop}) giving rise to the same conclusions -- the "background loop"
takes values into the dual torus to the compactified part of the
target-space. This is not much more than we had in sect.~\ref{ss:fermi},
and it is enough to compute the one-loop temperature partition function of a
superstring in NSR formalism \cite{AtWi}.
We see again, that the "vacuum expectation values" $a_\mu^{(k)}$ take
values in dual torus and, in complete analogy with the particle case,
they should be integer for the bosons, half-integer for the
fermions and generic point of a dual torus corresponds to anyons. So, in
order to compute the partition function one could "insert by hands" the
extra "${\bf Z}_2$ factor" for each fermionic sector propagating along the
loop in "temperature direction".

This is still not quite satisfactory prescription. On the other hand, setting
both $\langle\chi\Psi\rangle $ and $\langle\bar\chi\bar\Psi\rangle$
nonvanishing on the world-sheet one sees that the quartic term in
(\ref{fstring}) becomes also nonzero. Moreover, normalization (\ref{norm})
requires that the contribution of the quartic terms to the action is
proportional to the {\em inverse} powers $\beta^{-2}$ of the radius
$\beta$. It means that the string path integral sum over windings
respects duality
$\beta\leftrightarrow{\beta^{-1}}$ already {\em before} integrating
over two-dimensional geometries. Thus, we see that the properties
of NSR string path integral in presence of compact dimensions deserve
further careful analysis, in particular the questions of fermionic
condensates and background Wilson loops are not yet fully understood.

\section{Conclusion}

In this note we have considered the first-quantized formulation of the
temperature partition functions for the bosonic and fermionic particles and
a generalization of this path-integral representation to the case of
fermionic string. We have seen that the ensemble of particles is generated
by wrappings of the world-lines (world-sheets) over the compact direction in
space-time. It was shown that in the case of fermions additional phase factor
arises due to existence of nontrivial background Wilson loops of
${\bf Z}_2$ gauge fields and they can be interpreted even as vacuum
condensates of the world-line or world-sheet fermions.

The first-quantized integrals of this kind may be also applied for the
computations of string path integrals in the presence of D-branes (see,
for example, \cite{AMSS} and references therein). In
particular, it is easy to see that the partition function of strings in the
background of D-branes does not differ too much from the old propagator
computations \cite{CMNP,FM2} and leads, say, for
$N$-branes lead to an integral with $N$ boundaries of the typical form
\be
\int \prod_{i=1}^N dX_i~e^{-\sum_{i<j}(X_i-X_j)^2}\dots
\ee
as an "eigenvalue" matrix integral. It would be interesting to study
fermionic string path integrals of this sort, where the main problems are
expected in application of the NSR formalism (before GSO projection!) for the
computations with D-branes. One may expect all problems considered in
this note together with imposing of boundary conditions on the NSR fermionic
fields on the surfaces of D-branes.

Finally, let us point out that we have demonstrated in this note that the
"thermodynamic" computations in the first-quantized formalism for the
fermions, involving nontrivial effects with closed trajectories, do not
differ too much from the old computations in spirit of \cite{FM1,FM3}.
It reminds the similarity between the two interesting physical outcomes of the
first-quantized string computations \cite{FM2,AtWi}, related
to the exponential growth of the density of massive string states. In the "field
theory" part this is a modification of causality behavior of the
string theory Green functions \cite{FM2} while in "thermodynamic part"
it is related with the famous Hagedorn phase transition (see \cite{AtWi}
and references therein). Both effects are different sides of the same coin
and our arguments from sect.~\ref{ss:string} may serve as an extra sign
in favour of appearance of all "thermodynamic" effects directly from the NSR
string path integral.

\section*{Acknowledgements}

I am indebted to V.Fainberg, A.Losev, Yu.Makeenko, A.Rosly, I.Tyutin and
M.Voloshin
for interesting and important discussions. The work was partially supported
by RFBR grant No. 99-02-16122, INTAS grant No. 99-0590, CRDF grant No.
RP1-2102 ($\# $6531), NATO Collaborative Linkage Grant PST.CLG.977361 and
grant for the support of scientific schools No. 00-15-96566.

\end{document}